# Diffusion MRI with free gradient waveforms on a high-performance gradient system: Probing restriction and exchange in the human brain

Arthur Chakwizira[1], Ante Zhu[2], Thomas Foo[2], Carl-Fredrik Westin[3], Filip Szczepankiewicz[1], and Markus Nilsson[4]

1. Medical Radiation Physics, Clinical Sciences Lund, Lund University, Lund, Sweden
2. GE Research, Niskayuna, New York, USA
3. Department of Radiology, Brigham and Women's Hospital, Harvard Medical School, Boston, MA, United States
4. Department of Clinical Sciences Lund, Radiology, Lund University, Lund, Sweden

**Corresponding author:**

Arthur Chakwizira

Department of Medical Radiation Physics, Lund University, Skåne University Hospital,

SE-22185 Lund, Sweden

Email address: arthur.chakwizira@med.lu.se

## Abstract

The dependence of the diffusion MRI signal on the diffusion time carries signatures of restricted diffusion and exchange. Here we seek to highlight these signatures in the human brain by performing experiments using free gradient waveforms that are selectively sensitive to the two effects. We examine six healthy volunteers using both strong and ultra-strong gradients (80, 200 and 300 mT/m). In an experiment featuring a large set of gradient waveforms with different sensitivities to restricted diffusion and exchange (150 samples), our results reveal unique time-dependence signatures in grey and white matter, where the former is characterised by both restricted diffusion and exchange and the latter predominantly exhibits restricted diffusion. Furthermore, we show that gradient waveforms with independently varying sensitivities to restricted diffusion and exchange can be used to map exchange in the human brain. We consistently find that exchange in grey matter is at least twice as fast as in white matter, across all subjects and all gradient strengths. The shortest exchange times observed in this study were in the cerebellar cortex (115 ms). We also assess the feasibility of future clinical applications of the method used in this work, where we find that the grey-white matter exchange contrast obtained with a 25-minute 300 mT/m protocol is preserved by a 4-minute 300 mT/m and a 10-minute 80 mT/m protocol. Our work underlines the utility of free waveforms for detecting time-dependence signatures due to restricted diffusion and exchange *in vivo*, which may potentially serve as a tool for studying diseased tissue.

# 1 Introduction

Time-dependent diffusion MRI enables non-invasive investigation of the microstructure of biological tissue, yielding indices sensitive to cell sizes (water restriction) and membrane permeability (water exchange)(Callaghan et al., 1991; Kärger, 1985; Reynaud, 2017). Measurement of these tissue properties is important for the characterisation of pathology and monitoring of treatment response (Jiang et al., 2016; Ruggiero et al., 2018; Volles et al., 2001). Cell size measurements have been used in the analysis of blood in conditions such as anaemia and cancer (Evans and Jehle, 1991; Montagnana and Danese, 2016; Price-Jones, 1922; Xu et al., 2021) and to differentiate between cancer and normal cells (Shashni et al., 2018). Alterations in permeability signify compromised membrane integrity which is associated with conditions such as hydrocephalus, neuromyelitis optica, cancer and Parkinson's disease (Papadopoulos and Verkman, 2012; Sønderby et al., 2014; Verkman et al., 2017, 2008).

Estimation of restriction and exchange with diffusion MRI evidently holds value. However, accurate quantification of these processes is challenging and estimations found in literature have been inconsistent (Jelescu et al., 2022; Khateri et al., 2022; Lee et al., 2018; Novikov, 2021; Paquette et al., 2020; Veraart et al., 2020). Typical examples include the gross overestimation of axonal diameters (Alexander et al., 2010; Lee et al., 2020a; Xu et al., 2014; Zhang et al., 2011) and the inconsistent exchange times reported in grey matter (Jelescu et al., 2020; Olesen et al., 2022). The challenge is attributable to two problems: the reliance on inaccurate models and the limited information in the diffusion-weighted signal. To some degree, the former problem is circumvented by opting for parsimonious signal representations that make fewer assumptions about the source of the signal (Novikov et al., 2018). Examples include exchange estimation using cumulants of the phase distribution (Ning et al., 2018) and size estimation using the Gaussian phase approximation (Nilsson et al., 2017). The latter problem stems from a combination of suboptimal experimental designs and hardware constraints (Chakwizira et al., 2022; Nilsson et al., 2017; Paquette et al., 2020; Yadav et al., 2010). An example of a suboptimal experimental design is the probing of time dependence by varying the temporal separation of two diffusion encoding pulses in a single diffusion encoding (SDE) experiment, as it conflates the effects of restricted diffusion and exchange (Chakwizira et al., 2022; Nilsson et al., 2009; Olesen et al., 2022). The primary hardware constraint pertains to the gradient system in general and the gradient amplitude in particular. Here, the upper limit of clinical hardware has been pushed by the recent development of MRI scanners with ultra-strong gradient systems (Fan et al., 2022; Foo et al., 2020; Huang et al., 2020). Prominent examples include the 'Connectome scanners' which offer maximum gradient amplitudes of 300 mT/m at slew rates of 200 T/m/s (Fan et al., 2022; Huang et al., 2021; Setsompop et al., 2013) and the MAGNUS MRI scanner which delivers a maximum gradient amplitude of 200 mT/m at a slew rate of 500 T/m/s (Foo et al., 2020). The MAGNUS MRI scanner has recently been upgraded to a maximum gradient strength of 300 mT/m at a slew rate of 750 T/m/s. These technological advancements facilitate improved microstructure mapping, especially when combined with diffusion MRI frameworks that: (1) incorporate the effects of both restricted diffusion and exchange and (2) leverage experimental designs that disentangle effects of restricted diffusion and exchange.

We have in a previous work proposed a theoretical framework and accompanying experimental design for accurate mapping of restricted diffusion and exchange (Chakwizira et al., 2022). This framework extends the conventional one-dimensional diffusion time and defines, for a given gradient waveform, two parameters: the restriction weighting ($V_\omega$) and the exchange weighting (Γ) (Chakwizira et al., 2022). By independently varying these parameters using arbitrarily modulated gradient waveforms, the individual contributions of restricted diffusion and exchange to the signal can be disentangled, thereby enabling a more specific means to probe tissue microstructure. Consequently, compartment size and exchange rate indices estimated with this method exhibit less crosstalk than those obtained by just varying the gradient pulse separation in an SDE design. Estimates obtained using arbitrary gradients are also more precise than those obtained with pulsed gradients because arbitrary gradients provide a much broader range of restriction- and exchange-weightings. However, the range is still limited by the gradient system and human physiology (Chronik and Rutt, 2001; Jones et al., 2018; Setsompop et al., 2013). The gradient limit is primarily because the gradient waveforms used to probe restriction and exchange are oscillatory, which limits the maximum b-value that can be achieved given a certain total duration and maximum gradient amplitude (Szczepankiewicz et al., 2021). However, an efficient probe of exchange requires high b-values because the effect manifests in the fourth order cumulant of the phase distribution. The solution to this problem is to use ultra-strong gradients that yield high b-values for oscillatory waveforms in a relatively short time. Human physiology places a limit on the maximum slew rate attainable with ultra-strong gradients because high slew rates induce peripheral nerve stimulation (PNS) (Chronik and Rutt, 2001; Jones et al., 2018). This problem is partly addressed by the use of smaller head-only gradient coils that induce lower PNS because the shoulders and torso are outside the main gradient field (Setsompop et al., 2013).

The aims of this work were twofold. First, to investigate how the diffusion-weighted signal in the human brain varies with the restriction- and exchange-weighting parameters $V_\omega$ and Γ. Second, to study the potential benefit of using ultra-strong gradients for the joint estimation of restriction- and exchange-related parameters. To this end, we apply our previously proposed restriction-exchange framework to *in vivo* studies of the human brain using both strong gradients on a whole-body system (80 mT/m) and ultra-strong gradients on the MAGNUS system (200 and 300 mT/m). The ultra-strong gradients enabled the execution of what we denote a "discovery experiment" using the broadest set of unique gradient waveforms used to probe exchange in living tissue to date (150 different waveforms). Such a large collection was necessary to exhaustively sample the set of possible restriction- and exchange-weightings and test the applicability of the framework. Moreover, a faster protocol featuring fewer waveforms was also executed at three different levels of the maximum gradient amplitude (80, 200 and 300 mT/m). The waveforms in this experiment were designed to be selectively sensitive to restricted diffusion and exchange, thus facilitating independent parameter estimates. Our results indicate an unprecedented ability to disentangle restriction- and exchange-driven signal contrasts in the human brain, suggesting that the parameters $V_\omega$ and $\Gamma$ indeed capture independent time-dependence mechanisms. The results also confirmed that probing restricted diffusion demands ultra-strong gradients while exchange can be reliably studied even at 80 mT/m. Feasible exchange estimates were obtained especially in grey matter, which suggests the potential of the method for the assessment of cortical pathology.

## 2 Theory

The effects of restriction and exchange on the diffusion-weighted signal, $S$, are captured by the following signal representation (Chakwizira et al., 2022; Nilsson et al., 2017; Ning et al., 2018)

$$\ln(S/S_0) \approx -b \cdot \left[E_{\beta_0} + V_\omega E_{\beta_2}\right] + \frac{1}{2}b^2 \cdot \left[V_{\beta_0} + 2V_\omega C_{\beta_0\beta_2} + V_\omega^2 V_{\beta_2}\right] \cdot (1 - k\Gamma), \quad (1)$$

where three parameters describe the experiment ($b$, $V_\omega$ and $\Gamma$), six parameters describe the diffusion process ($E_{\beta_0}$, $E_{\beta_2}$, $V_{\beta_0}$, $C_{\beta_0\beta_2}$, $V_{\beta_2}$, and $k$) and one the non-diffusion-weighted signal ($S_0$). The core idea of the framework is that effects of restriction and exchange can be disentangled by using multiple diffusion encoding gradient waveforms defined so that combinations of $V_\omega$ and $\Gamma$ vary in a non-colinear manner.

Among the three experiment-related parameters, the b-value ($b$) is the strength of the diffusion weighting. It is defined through $b = \int_0^T q^2(t)\,dt$ where $q(t)$ is the dephasing q-vector, $q(t) = \gamma \int_0^t g(\tau)d\tau$, and $\gamma$ is the gyromagnetic ratio. The second parameter ($V_\omega$) quantifies the strength of restriction encoding and is defined as (Nilsson et al., 2017)

$$V_\omega = \frac{1}{2\pi b} \int_{-\infty}^{\infty} |q(\omega)|^2 \omega^2 \, d\omega = \frac{\gamma^2}{b} \int_0^T g^2(t)dt, \tag{2}$$

where $q(\omega)$ is the Fourier transform of $q(t)$. The third and final parameter ($\Gamma$) quantifies the exchange weighting and is given by (Ning et al., 2018)

$$\Gamma = 2 \int_0^T t \, \widetilde{q_4}(t) \, dt, \tag{3}$$

where $\widetilde{q_4}(t) = q_4(t)/b^2$ and $q_4$ is the fourth-order autocorrelation function of the dephasing q-vector defined as $q_4(t) = \int_0^T q^2(t')q^2(t'+t)dt'$.

Among the diffusion-related parameters, $E_{\beta_0} = \langle \beta_0 \rangle$ is the mean time-independent diffusivity, $E_{\beta_2} = \langle \beta_2 \rangle$ is the average restriction coefficient, which for diffusion restricted in a set of compartments has an expected value given by $E_{\beta_2} = f_R \langle cd^4/D_0 \rangle$ where $f_R$ is the fraction of restricted compartments in the voxel, $d$ is the compartment diameter, $D_0$ is the bulk diffusivity and $c$ is a constant. Furthermore, $V_{\beta_0}$ and $V_{\beta_2}$ are the variances in $\beta_0$ and $\beta_2$, $C_{\beta_0\beta_2}$ is the covariance between $\beta_0$ and $\beta_2$. Finally, $k$ is the exchange rate, which for a two-compartment system has an expected value of $k = k_{12} + k_{21}$ where $k_{12}$ and $k_{21}$ are the forward (e.g., intra-to-extra cellular) and reverse exchange rates, respectively.

## 3 Methods

### 3.1 Protocol design

Five different acquisition protocols were designed and applied in this study. Two of these were 'discovery protocols', one defined with a maximum amplitude of 200 mT/m and another with 300 mT/m, aimed at investigating whether the change in signal in tissue obtained by varying the parameters $V_\omega$ and $\Gamma$ followed the relationship predicted by Eq. 1. The remaining three protocols were so-called 'application protocols,' designed to facilitate estimation of the microstructural parameters in Eq. 1. These three protocols were designed for maximum gradient amplitudes of 80, 200, and 300 mT/m to study how the maximum amplitude impacted the quality of parameter estimates. All five protocols are described below and in Table 1.

*Design of the discovery protocols*

The discovery protocols comprised a set of 150 unique free gradient waveforms designed to exhibit different values of $V_\omega$ and $\Gamma$. The waveforms were designed using the same procedure described in ref (Chakwizira et al., 2022) and generated to utilize a maximum gradient of 200 mT/m and 220 T/m/s for the first protocol and 300 mT/m and 195 T/m/s for the second. These slew rates were lower than the system capability but were chosen to avoid exceeding PNS thresholds. The total diffusion encoding gradient duration was limited to 100 ms (to ensure a TE shorter than 120 ms for sufficient SNR), and the minimum pause duration was set to 5.2 ms (to accommodate the refocusing RF pulse). Symmetry about the refocusing pulse was enforced and a b-value of at least 4 ms/µm² at the maximum gradient amplitude was required.

Figure 1A shows the 300 mT/m discovery protocol where the 150 grey dots correspond to 150 unique waveforms that spanned the space of restriction- and exchange-weightings. Some example waveforms are highlighted along the boundary where short-duration SDE-like waveforms provide the weakest exchange weighting and double diffusion encoding (DDE)-like waveforms deliver strongest exchange weighting. SDE-like pulses with short spacing but long duration give weak restriction weighting while oscillatory waveforms generally yield strong restriction weighting. The 200 mT/m discovery protocol is depicted in Fig. A1A of the supplementary material.

*Design of the 200 and 300 mT/m application protocols*

Application protocols were developed to efficiently probe restricted diffusion and exchange. Each protocol featured twelve waveforms. Six of these had fixed restriction weighting but varying exchange weighting and the remaining six had fixed exchange weighting but varying restriction weighting. The waveforms were subjected to the same constraints as the corresponding discovery protocols and implemented on the same scanner.

The 300 mT/m application protocol is shown in Fig. 1B and 1C where the blue waveforms are variably sensitive to exchange (x-axis) and the red waveforms are variably sensitive to restricted diffusion (y-axis). The 200 mT/m application protocol is presented in Fig. A1B of the supplementary material.

*Design of the 80 mT/m application protocol*

A set of four free gradient waveforms was designed for a maximum gradient amplitude of 80 mT/m, maximum slew rate of 70 T/m/s, maximum total diffusion encoding gradient duration of 120 ms, minimum pause duration of 9 ms, symmetry about the refocusing pulse and a minimum b-value of 5 ms/µm² at the maximum gradient strength. The resulting protocol comprised two waveforms with varying restriction weighting and fixed exchange weighting and two waveforms with varying exchange weighting and fixed restriction weighting (Fig. A1C).

Note that while the ranges of $\Gamma$ for the 200 mT/m and 80 mT/m protocols are comparable, the range of $V_\omega$ for the 80 mT/m protocol is notably smaller than for the 200 mT/m protocol. This is because waveforms with strong restriction weighting are generally oscillatory and thus less efficient (Szczepankiewicz et al., 2021). The requirement of high b-values excluded such waveforms at the lower gradient strength.

## 3.2 I*n vivo* experiments

*Experiments using the 200 and 300 mT/m discovery protocols*

The discovery protocols were implemented on a GE SIGNA 3.0 T MR750 scanner equipped with a 42-cm diameter head-only gradient coil (MAGNUS). The system operates at a maximum gradient of 200 mT/m and 500 T/m/s with 1 MVA gradient driver, and a maximum gradient of 300 mT/m and 750 T/m/s with 2 MVA gradient driver. The 200 mT/m protocol was applied on a single healthy volunteer using the pulse sequence parameters TE = 115 ms, TR = 3 s, resolution of $2 \times 2 \times 5$ mm³ and 16 slices. All 150 waveforms were applied using three b-values: 0, 2 and 4 ms/µm² along a single direction (anterior-posterior). A single direction was chosen to invest the scan time into a large variety of waveforms instead of rotations of the same waveforms. Gradients were directed along the anterior-posterior axis to attenuate the major white matter tracts of the cerebellum to enhance the grey-white matter contrast. Using the 300 mT/m discovery protocol, five healthy volunteers were scanned using all the same sequence parameters as for the 200 mT/m protocol, except for the echo time which was 112 ms. The scan time for the discovery protocols was 15 minutes each.

*Experiments using the 200 and 300 mT/m application protocols*

The 200 mT/m application protocol was deployed on the same healthy volunteer and using the same pulse sequence parameters as the 200 mT/m discovery protocol. Each waveform in the application protocol was applied in 10 directions per non-zero b-value using the b-values 0, 1, 2, 3 and 4 ms/µm$^2$. The 300 mT/m application protocol was applied on the same five healthy volunteers and with the same pulse sequence parameters as the 300 mT/m discovery protocol. Waveform rotation and scaling was the same as for the 200 mT/m application protocol. The scan time for the application protocols was 25 minutes each.

*Experiments using the 80 mT/m application protocol*

A 3T MAGNETOM Prisma (Siemens Healthcare, Germany) with a prototype pulse sequence that accommodates user-defined gradient waveforms (Szczepankiewicz et al., 2019) was used to execute the 80 mT/m application protocol in a healthy volunteer. All four gradient waveforms were performed in 42 directions using b-values of 0, 0.25, 1.75, 3.5, and 5 ms/µm$^2$. Imaging parameters were TE = 136 ms, TR = 3.5 s, resolution of $2 \times 2 \times 5$ mm$^3$, 20 slices, strong fat suppression, and a total acquisition time of 10 minutes.

Table 1: Protocol parameters used in the *in vivo* experiments. The 12-minute and 4-minute protocols are subsamples of the 300 mT/m application protocol.

| Protocol | Number of waveforms | b-values [ms/μm²] | Number of directions | Γ interval [ms] | $V_\omega$ interval [s⁻²] |
|---|---|---|---|---|---|
| 300 mT/m discovery | 150 | 0, 2, 4 | 1, 1, 1 | 5 – 40 | 1400 – 54000 |
| 300 mT/m application | 12 | 0, 1, 2, 3, 4 | 1, 10, 10, 10, 10 | 5 – 40 | 1900 – 54000 |
| 200 mT/m discovery | 150 | 0, 2, 4 | 1, 1, 1 | 6 – 37 | 1000 – 22000 |
| 200 mT/m application | 12 | 0, 1, 2, 3, 4 | 1, 10, 10, 10, 10 | 6 – 36 | 1000 – 21000 |
| 80 mT/m application | 4 | 0, 0.25, 1.75, 3.5, 5 | 1, 6, 6, 12, 16 | 11 – 35 | 750 – 3400 |
| 12-minute 300 mT/m application | 6 | 0, 1, 2, 3, 4 | 1, 10, 10, 10, 10 | 5 – 40 | 1900 – 54000 |
| 4-minute 300 mT/m application | 4 | 0, 1, 3, 4 | 1, 10, 10, 10 | 5 – 40 | 1900 – 54000 |

*Ethics statement*

All human MRI scans were in accordance with local Institutional Review Board-approved protocols. All research MRI scans were performed after obtaining written informed consent from each subject.

## 3.3 Numerical simulations

To illustrate the performance of the approach in microstructural environments with controlled levels of restricted diffusion and exchange, Monte Carlo simulations were performed in a substrate of equally sized, regularly packed cylinders. The cylinder diameters were 2, 4, 6, 8, and 10 µm in separate simulations. The exchange rate was set to zero in these cases. Exchange was then introduced at rates of 4, 8, 12, 16, and 20 s$^{-1}$ at a fixed diameter of 4 µm. In each case, signals were generated using all 150 waveforms from the 300 mT/m discovery protocol at a b-value of 4 ms/µm² and the 12 waveforms from the 300 mT/m application protocol at b-values of 0, 1, 2, 3, and 4 ms/µm². In all simulations, gradients were applied perpendicular to the cylinder axis. Other simulation parameters were a bulk diffusivity of 2 µm²/ms, 10$^5$ particles, intracellular and extracellular fractions of 0.5, and a simulation time-step of 0.5 µs. The simulations followed the same procedure as described in detail by Chakwizira et al. (Chakwizira et al., 2022).

## 3.4 Data analysis

*Simulations*

Simulated data for the 300 mT/m discovery protocol were analysed by visualising the signals at a fixed b-value as a function of both $V_\omega$ and Γ. The simulated signals were compared to predicted signals obtained by fitting Eq. 1. Throughout this work, the parameters $C_{\beta_0\beta_2}$ and $V_{\beta_2}$ were constrained to zero because the variance term was found to be dominated by $V_{\beta_0}$ for all protocols considered in this work (data not shown). Our previous work highlighted that the approximation involved in the definition of Γ could induce bias in the analysis (Chakwizira et al., 2022). To examine this potential bias, we also fitted a more complete version of Eq. 1 given by (Ning et al., 2018)

$$\ln(S/S_0) \approx -b \cdot [E_{\beta_0} + V_\omega E_{\beta_2}] + \frac{1}{2} b^2 \cdot V \cdot h(k), \quad (4)$$

where $V = V_{\beta_0}$, $h(k) = 2 \int_0^T \widetilde{q_4}(t) \cdot \exp(-kt) \, dt$. An analysis of the fit residuals was then performed to evaluate the extent to which the signal representation explains the variation in the measured signals.

Simulated data for the 300 mT/m application protocol were analyzed by visualizing signal-vs-b curves for all waveforms in the protocol. Furthermore, both equations 1 and 4 were fitted to the signals to obtain parameter estimates using the non-linear least-squares solver *lsqnonlin* in MATLAB (The MathWorks, Natick, MA, R2022a).

*In vivo experiments*

Diffusion-weighted images from both the discovery and application protocols were post-processed in two steps. First, the images were denoised using the Marchenko Pastur MPPCA algorithm (Veraart et al., 2016). Second, motion and eddy-current correction was performed following a registration of the images to extrapolated references using ElastiX (Klein et al., 2010). Extrapolation-based references have been shown to give more accurate registration of high b-value data as were acquired in this work (Nilsson et al., 2015).

In vivo data from the discovery protocols were analysed in the same way as was done for the simulated data described above, with the measured signals being extracted from different regions of interest.

Data from the application protocols were analysed by inspecting the signal attenuation in different regions of interest. All regions of interest were drawn manually in each subject. Parameter estimates were obtained by the voxel-wise fitting of Eq. 4 to the powder-averaged signals using *lsqnonlin* in MATLAB (The MathWorks, Natick, MA, R2022a).

Of the five volunteers that were studied at 300 mT/m, one subject was excluded from this study on account of considerable motion artefacts that could not be corrected for. The study thus involved 6 subjects in total: 1 at 80 mT/m, 1 at 200 mT/m, and 4 at 300 mT/m.

*Subsampled application protocols*

The feasibility of the approach explored in this work for potential clinical use was assessed by performing the fit of Eq. 4 to two subsets of the 300 mT/m datasets: a 12-minute protocol obtained by reducing the number of waveforms from 12 to 6 and a 4-minute protocol obtained by additionally removing high-b-value acquisitions for the restriction-encoding waveforms and low-b-value acquisitions for the exchange-encoding waveforms. The parameters of the two protocols are presented in Table 1. Following the subsampling of the data, the same pre-processing steps and subsequent parameter estimation as described for the full dataset were performed.

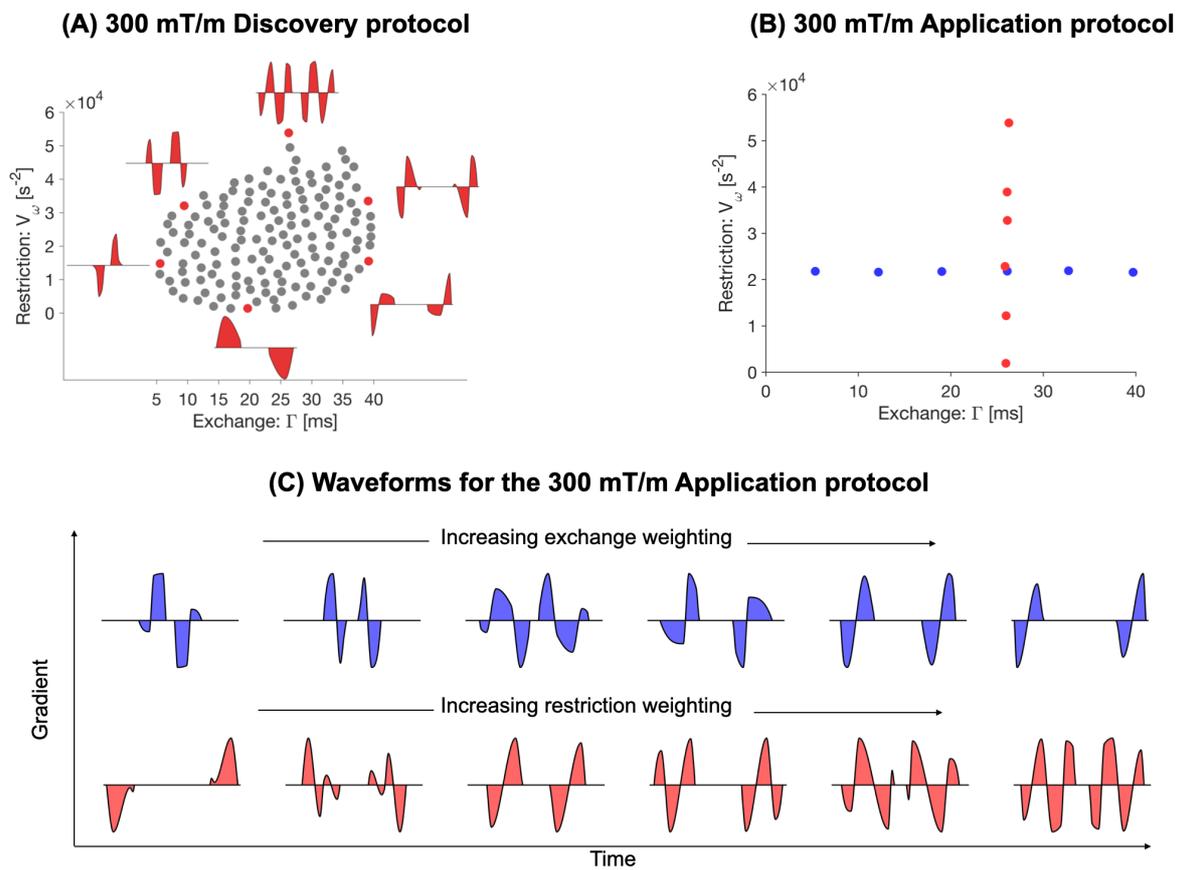

Figure 1: Free-waveform protocols used to probe restriction and exchange *in vivo* at 300 mT/m. Panel (A) shows the discovery protocol comprising 150 waveforms with different values of $V_\omega$ and $\Gamma$. All waveforms were generated under the constraints of having a maximum gradient strength of 300 mT/m, a slew rate of 195 T/m/s, a total duration of 100 ms, and a minimum b-value of 4 ms/µm$^2$ at the maximum gradient amplitude. Panel (B) shows the restriction-exchange weightings for the 300 mT/m application protocol. Panel (C) shows the 300 mT/m application protocol featuring 12 waveforms where 6 are exchange-encoding (similar $V_\omega$ but varying $\Gamma$) and the other 6 are restriction-encoding (similar $\Gamma$ but varying $V_\omega$). These waveforms were designed following the same approach as in (A). Waveforms in (A) and (C) were designed for the MAGNUS system (Foo et al., 2020), which delivers a maximum gradient strength of 300 mT/m and slew rate of 750 T/m/s.

# 4 Results

Synthetic signals generated using the 300 mT/m discovery protocol are shown in Fig. 2 where each point denotes the signal from each of the 150 waveforms at a b-value of 4 ms/µm². The general trend is that, in the absence of exchange, the signal variation takes place along the $V_\omega$-axis. Introducing exchange causes a signal variation along the Γ-axis. Thus, a signal variation along the Γ-axis indicates that exchange dominates the time-dependence, while a signal variation along the $V_\omega$-axis signifies that restricted diffusion is the more important mechanism. The fit residuals were small relative to the signals, indicating that the signal representation (Eq. 4) explains most of the variation in the signals. The non-zero residuals indicate that the waveforms do not fully respect the assumptions of Eq. 4 (Chakwizira et al., 2022).

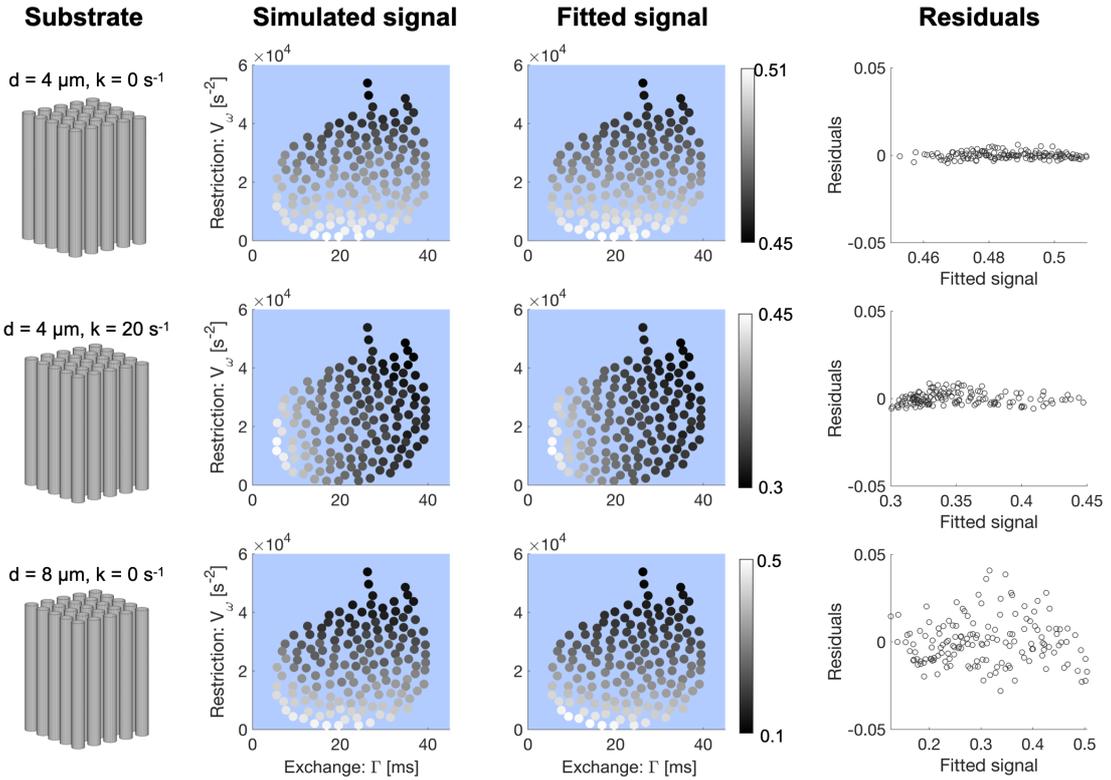

Figure 2: Simulated signals generated with the 300 mT/m discovery protocol. Simulations were performed in a substrate of regularly packed cylinders with a diameter of 4 µm without and with exchange (first two rows) and 8 µm without exchange (bottom row). The signals were obtained at a b-value of 4 ms/µm² and are plotted here as a function of $V_\omega$ and Γ. The brightness of each point represents the signal intensity. When the diameter is fixed and exchange is increased, the signal variation is found predominantly along the Γ-axis. In the absence of exchange, the signal variation is consistently along the $V_\omega$-axis.

Signals simulated with the 300 mT/m application protocol are shown in Fig. 3 as a function of both b-value and Γ and $V_\omega$. For small cylinder diameters (4 μm) in the absence of exchange, the exchange-encoding waveforms (blue) show negligible contrast compared to the restriction-encoding waveforms (red). Introducing exchange at 4 μm induces pronounced contrast in the exchange-encoding waveforms. At larger sizes (8 μm) in the absence of exchange, the restriction-encoding waveforms show greatest contrast. However, there is non-negligible contrast from the exchange-encoding waveforms as well, despite the lack of exchange in the simulations. This illustrates that large restrictions pose a challenge for the framework used in this work.

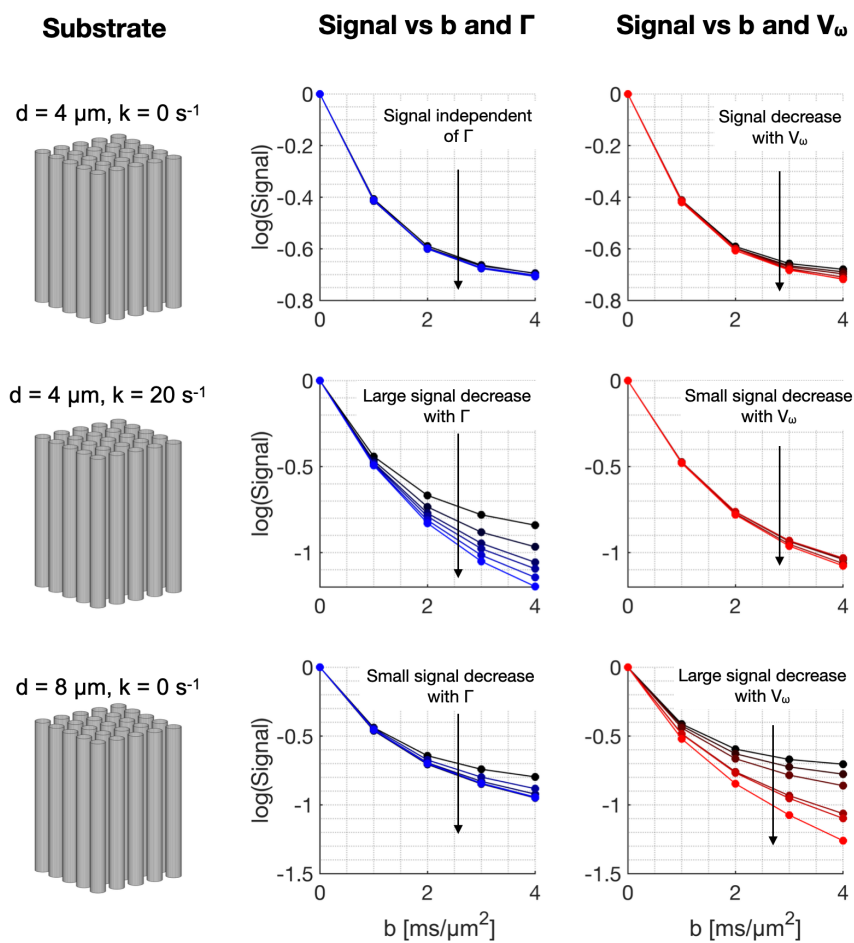

Figure 3. Simulated signal-vs-b curves using the 300 mT/m discovery protocol. Simulations were performed in substrates of parallel regularly packed cylinders of diameters 4 μm without and with exchange (first two rows) and 8 μm without exchange (bottom row). In the absence of exchange, time dependence at 4 μm manifests in $V_\omega$. In the presence of exchange at 4 μm, time dependence is dominated by exchange. At 8 μm and no exchange, time dependence manifests mainly in $V_\omega$ but is also seen to a lesser degree in Γ.

Figure 4 shows estimates of the restriction coefficient and exchange rate obtained by fitting Eq. 1 and Eq. 4 to the signals in Fig. 3. The restriction-coefficient estimates are generally in good agreement with expectations, regardless of the underlying exchange rate. Exchange estimates obtained by fitting both Eq. 1 (blue) and Eq. 4 (red) correlate well with the ground truth. The estimates exhibit bias that increases with the underlying exchange rate, but this bias is smaller when using Eq. 4 than Eq. 1. Exchange estimates are independent of the underlying size up to about 6 μm above which the exchange estimates depend on size.

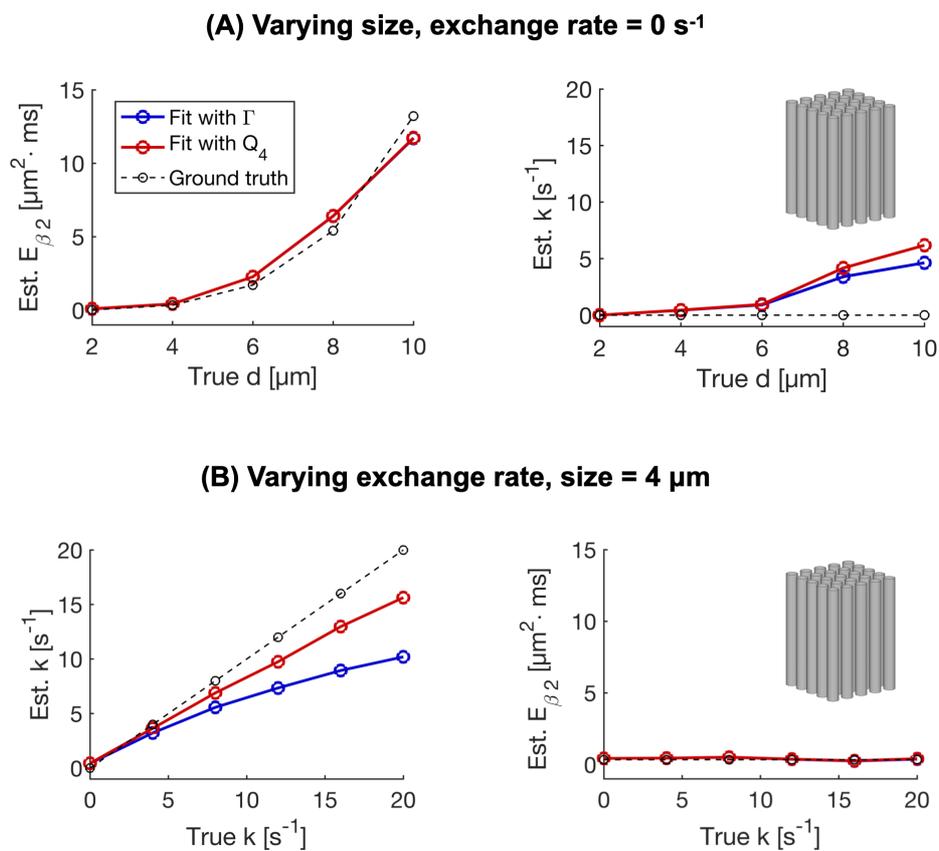

Figure 4: Estimates of the restriction coefficient and exchange rate obtained with the 300 mT/m application protocol. Simulations were run at zero exchange with sizes between 2 and 10 μm (panel A) and a fixed size of 4 μm with exchange rates between 0 and 20 $s^{-1}$ (panel B). The figure shows that for the simulation scenarios considered here, the restriction coefficient estimates generally agree with ground truth and are independent of the underlying exchange rate. Exchange estimates correlate well with ground truth but have a negative bias at higher exchange rates. The bias is smaller when using Eq. 4 ("Fit with Q4") than Eq. 1 ("Fit with Γ"). Exchange estimates are independent of the underlying size up to a size of 6 μm.

Signals intensities obtained *in vivo* with the 300 mT/m discovery protocol in volunteer 1 are shown in Fig. 5 for the cerebellar cortex and internal capsule. The measured signal refers to the average signals in the ROI at a fixed b-value of 4 ms/μm². Signals predicted by fitting the representation in Eq. 4 are shown alongside the measured signals. There is an evident trend where the time dependence (signal variation) appears to be along both the exchange and restriction axes in the cerebellar cortex (grey matter) and predominantly along the restriction-axis in the internal capsule (white matter). Figure 5 also displays the correlation between the residuals from the fitting described above and the measured signals. The residuals are of the same order of magnitude as those observed when fitting to noiseless simulated signals (Fig. 2).

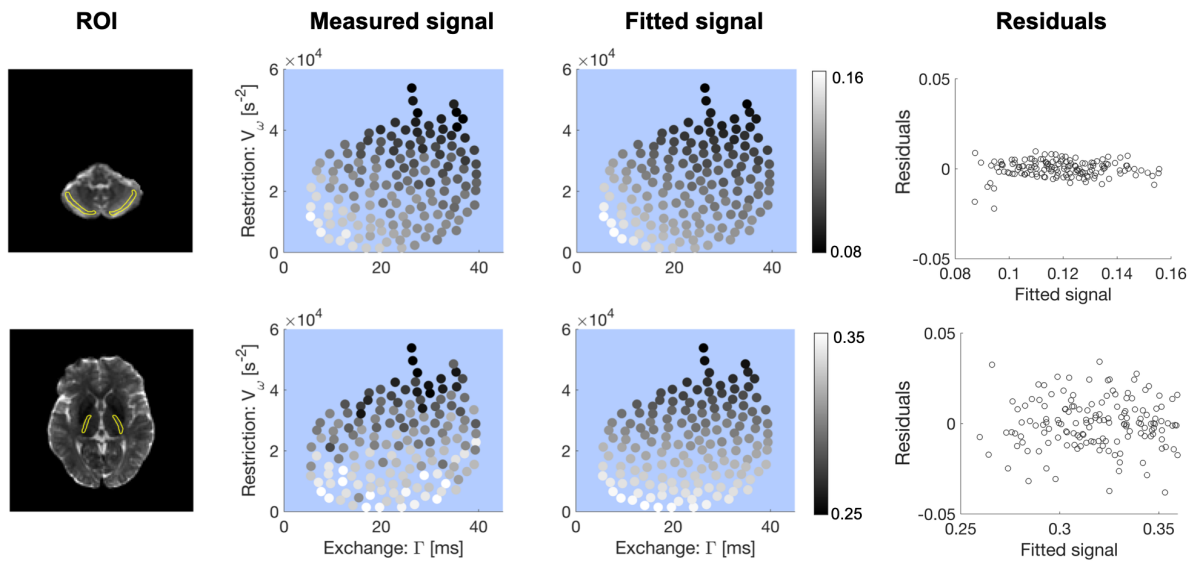

Figure 5: Contrast obtained by varying restriction- and exchange-weighting *in vivo*. Results are shown in two ROIs: one in the cerebellar cortex (top row) and anther in the internal capsule (bottom row). The measured signal is the average signal in each ROI for all 150 waveforms in the discovery protocol at a fixed b-value of 4 ms/μm², scaled by the non-diffusion-weighed signal. Darker dots mean lower signal while brighter ones mean higher signal. The fitted signal is the result of fitting Eq. 4 to the measured signals and the fit residuals from that fitting are shown alongside. The results shown suggests that time-dependence in grey matter is driven by both exchange and restriction (dots have a bright-dark trend along the diagonal) while time-dependence in white matter is driven by restriction (dots have a bright-dark trend along the y-axis).

Signal-vs-b curves for the 300 mT/m application protocol in volunteer 3 are shown in Fig. 6 in three brain regions: the cerebellar cortex, the internal capsule, and the thalamus. In the cerebellar cortex, exchange and restricted diffusion seem to be competing in driving the time-dependence. The same is true of the thalamus where both restriction and exchange appear to play an important role. The trend is different in the internal capsule where restriction appears to dominate. Overall, Fig. 6 suggests that time dependence in grey matter is driven by both restricted diffusion and exchange while restriction is more important than exchange in white matter. Note that the signal-vs-b curves in each group in Fig. 6 seem to share a common initial slope and only diverge at higher b-values. This indicates that high b-values are a prerequisite for highlighting the time dependence reported in this work.

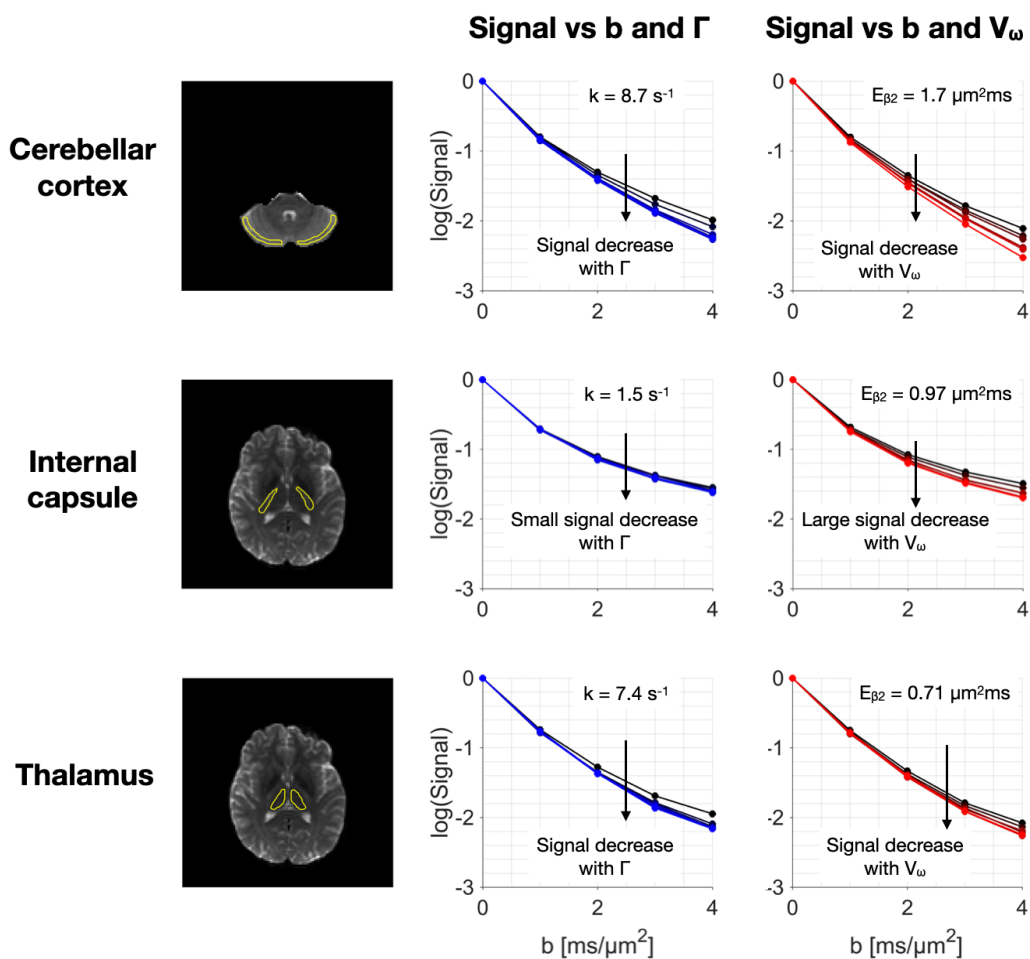

Figure 6: Signal-vs-b curves in different brain regions. The first column shows ROIs in the cerebellar cortex (grey matter), internal capsule (white matter) and thalamus (deep grey matter). The second and third columns show the average signal in the ROI as a function of b-value for the six exchange-encoding waveforms (second column) and the six restriction-encoding waveforms (third column) in the 300 mT/m application protocol (Fig. 1C). In the cerebellar cortex, time dependence is influenced to by both exchange and restriction. In the internal capsule, time dependence appears to be dominated by restriction. The thalamus exhibits the same trend as the cerebellum where both exchange and restriction seem to drive the time dependence. Overall, the figure suggests that restriction dominates time-dependence in white matter while exchange and restriction are important in grey matter.

Parameter maps obtained from fitting the representation in Eq. 4 to powder-averaged signals acquired with the 80, 200 and 300 mT/m application protocols are presented in Fig. 7 for the cerebrum. Except for high values in the corticospinal tract, the restriction maps generally show little contrast in all subjects. In the exchange maps, the grey-white matter contrast is plentiful, with clear delineation of subcortical structures such as the head of caudate nucleus and the putamen. This is true of all subjects and at all gradient strengths. Exchange estimates in grey matter are higher than those in white matter by at least a factor of two, as summarised in Table 2. The table also shows a general increase in the estimated exchange rate with increasing gradient strength.

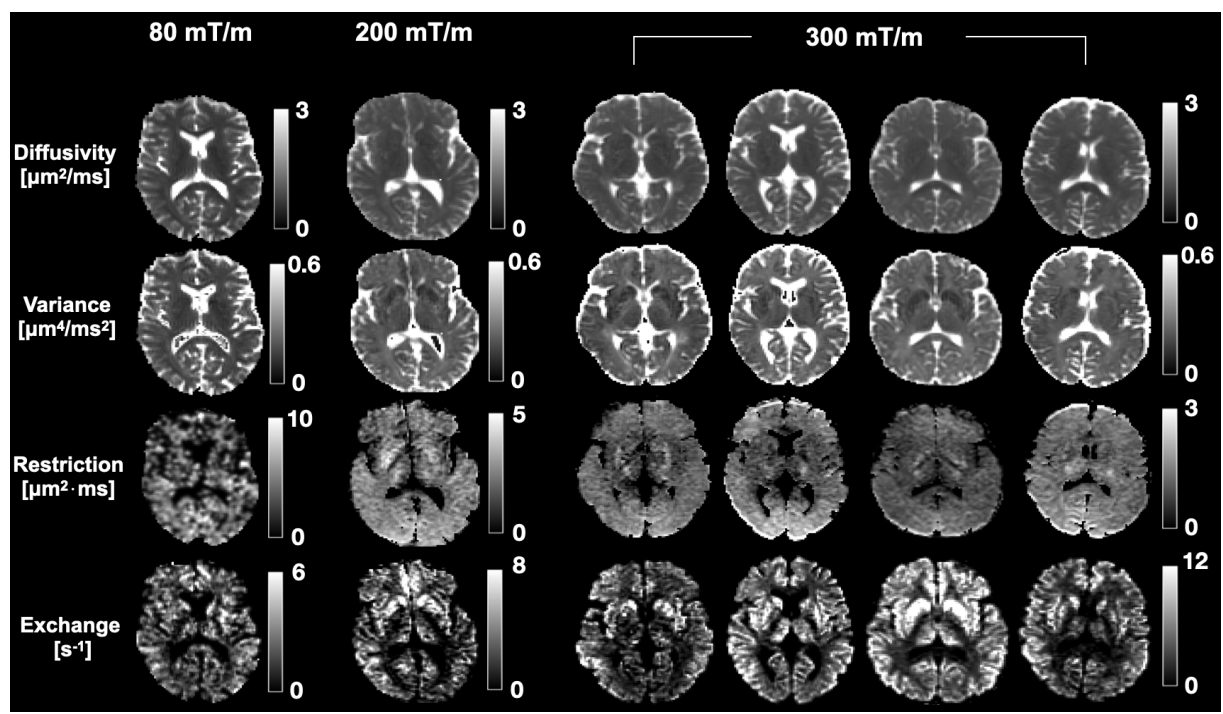

Figure 7: Parameter estimates obtained with the 80, 200 and 300 mT/m application protocols in the cerebrum for all scanned subjects. The maps are the result of fitting the signal representation in Eq. 4 to powder-averaged signals. While the restriction maps are generally featureless, the exchange maps show plentiful contrast, both within and across subjects. The exchange maps display higher values in grey matter than in white matter. The 80 mT/m protocol reveals the same trends in exchange estimates as the 200 and 300 mT/m but with lower grey-white matter contrast.

Parameter maps from the cerebellum are shown in Fig. 8. The restriction coefficient also shows little contrast between cerebellar cortex and white matter. However, the restriction maps are hyperintense around the dentate nucleus which is a large cluster of neurons located in the cerebellar white matter. The contrast in the exchange maps from the cerebellum concurs with Fig. 7: notably faster exchange in grey matter compared to white matter. It is worth noting that the highest exchange rates observed in this study were in the cerebellar cortex (Table 2).

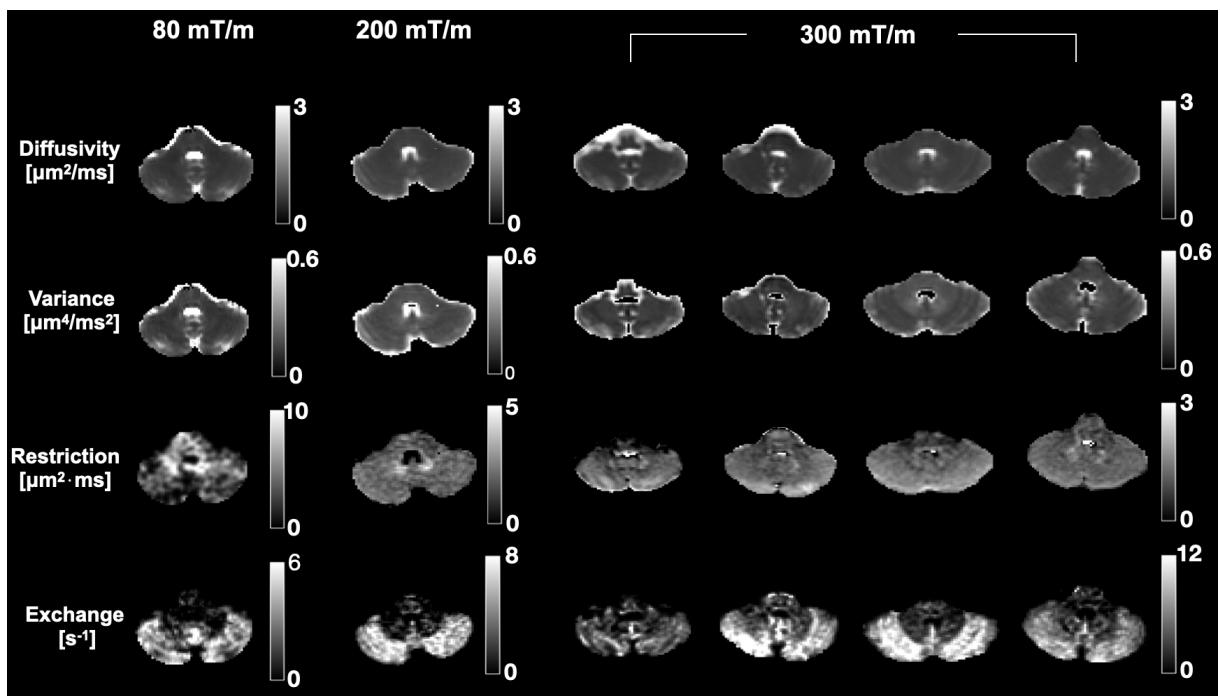

Figure 8: Parameter estimates obtained with the 80, 200, and 300 mT/m application protocols in the cerebellum for all scanned subjects. The maps were obtained by fitting Eq. 4 to powder-averaged signals. While the restriction maps are generally devoid of contrast, the exchange maps show a clear distinction between grey and white matter where the former shows notably faster exchange. There is visual correspondence between exchange maps from the 80 mT/m protocol and exchange maps at 200 and 300 mT/m.

Table 2: Exchange estimates obtained from fitting Eq. 4 to powder-averaged signals at 80, 200 and 300 mT/m. The table shows the mean (standard deviation) of the estimates in each ROI. The 300 mT/m results are shown for subject 3 as an illustration. The ROIs in frontal and parietal white matter can be found in Fig. A2 of the Appendix, while the other three are shown in Fig. 6. All ROIs were drawn manually in each subject. Exchange rates in grey matter are invariably higher than those in white matter. There is an increase in estimated exchange rates with increasing gradient strength.

| ROI | Exchange at 80 mT/m [$s^{-1}$] | Exchange at 200 mT/m [$s^{-1}$] | Exchange at 300 mT/m [$s^{-1}$] |
|---|---|---|---|
| Frontal white matter | 1.7 (0.9) | 1.9 (1.1) | 3.5 (1.6) |
| Parietal white matter | 1.6 (0.9) | 0.9 (0.9) | 2.9 (1.3) |
| Internal capsule | 0.5 (0.7) | 0.6 (0.8) | 1.5 (0.7) |
| Thalamus | 2.3 (1.8) | 3.1 (1.9) | 7.4 (2.6) |
| Cerebellar cortex | 3.5 (1.4) | 5.9 (1.2) | 8.7 (3.1) |

Figure 9 shows the parameter maps obtained using the 12-minute and 4-minute subsampled 300 mT/m application protocols. The results are shown for volunteer 2. The 12-minute protocol gives maps that are visually indistinguishable from those obtained with the 25-minute protocol (Fig. 7). A more striking result is that the 4-minute protocol, apart from a noticeable influence of noise, also provides maps that closely resemble those obtained with the 12-minute protocol. The exchange estimates are, however, lower than those obtained with the 12-minute protocol, which points towards multiexponential exchange that cannot be unambiguously captured by only two points. Note that the 4-minute protocol constitutes the minimal protocol needed to extract the parameters presented in this work and any further subsampling causes the fit to fail.

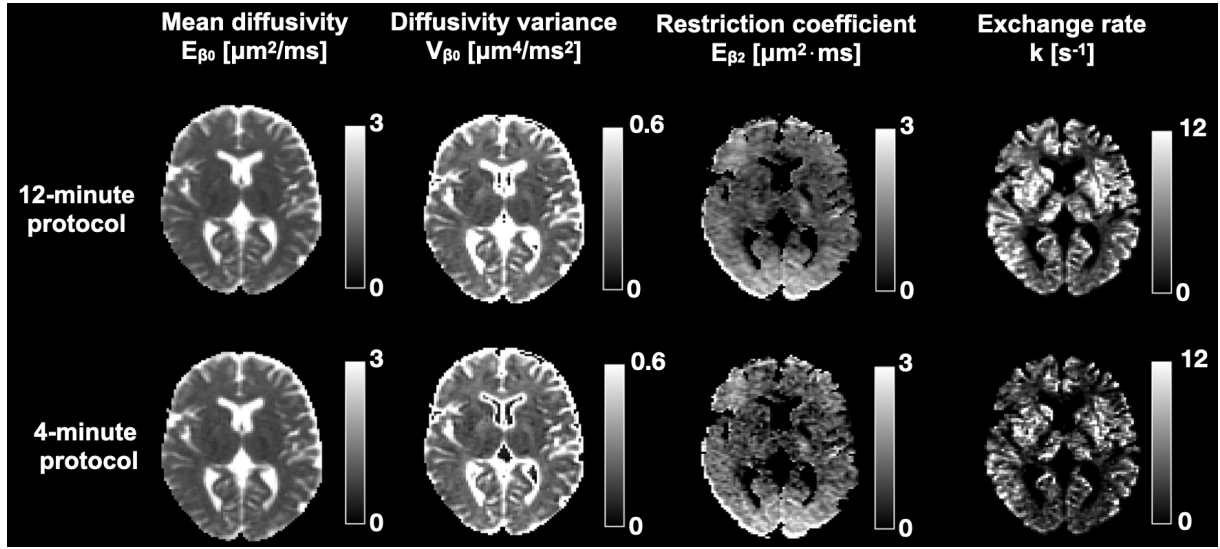

Figure 9: Parameter maps obtained after subsampling the 300 mT/m application protocol to 12 and 4 minutes. Results are shown in volunteer 2. The subsampling was done by reducing the number of waveforms and b-values. The 12-minute protocol gives maps essentially identical to the full 25-minute dataset. The 4-minute protocol shows higher sensitivity to noise as well as lower exchange rates, but preserves the contrasts seen in the longer protocols.

# 5 Discussion

This work showed that at total diffusion encoding times of up to 100 ms, diffusion time dependence in grey matter is influenced by both restriction and exchange while restriction dominates the time dependence in white matter. Moreover, exchange rates in grey matter were found to be at least twice as fast as those in white matter. This work also showed that exchange and restricted diffusion can be simultaneously probed at 300 mT/m in just 4 minutes, yielding parameter maps with the same contrasts as a 25-minute protocol. Furthermore, using free waveforms and the restriction-exchange framework applied in this study, exchange in the human brain can be probed at 80 mT/m, giving the same grey-white matter contrast as would be obtained at 300 mT/m.

The restriction- and exchange-weighting parameters used in this work sufficiently describe time dependence of the diffusion processes in the human brain. The observed trends, that the signals in grey and white matter respond distinctly to changes in $\Gamma$ and $V_\omega$ (Figs 5 and 6), indicate that these parameters expose the signatures imprinted by different microstructures on the diffusion-weighted signal. This claim is corroborated by numerical simulations (Figs 2-4) which showed distinct signal dependencies on Γ and $V_\omega$ when restriction and exchange were varied separately. The finding that exchange has a negligible influence on the time dependence of the signal from white matter suggests that, at these time scales, axons are effectively impermeable. This supports the modelling of axons as impermeable zero-radius sticks (Novikov et al., 2019) and aligns with previous findings using filter-exchange imaging (Nilsson et al., 2013a). The outcome that exchange in grey matter is non-negligible is also in agreement with recent studies which suggest fast exchange between neurites and soma or neurites and the extracellular space (Olesen et al., 2022). The discovery experiment is, to the best of our knowledge, the first time that restriction and exchange in living tissue have been probed using this broad a set of gradient waveforms with unique timing parameters. It is remarkable that the entire set of waveforms largely follows trends that are explicable based on knowledge of the cellular composition of the brain. Nevertheless, it is important to note that some studies have attributed observed time dependence in human white and grey matter to structural disorder rather than restriction and exchange (Fieremans et al., 2016; Lee et al., 2020b). Three comments can be made regarding the apparent inconsistency. First, the contemporary studies mentioned above employ standard SDE waveforms with finite fixed pulse widths and varying diffusion times which, as shown in previous work (Chakwizira et al., 2022), may inadvertently conflate the effects of restriction and exchange. As a result, the outcome of such studies cannot be easily compared with the findings of the present work in which we leverage waveforms that are designed to separate the effects of restriction and exchange. Second, while in the forward sense we claim that exchange in the tissue causes a signal variation with Γ, in the inverse sense we can only claim that a signal variation with Γ *possibly* reflects exchange in the tissue. That is, a signal variation with Γ may also result from structural disorder. Indeed, our previous work reported elevated exchange rates with increasing disorder in simulation substrates (Chakwizira et al., 2022). Third, and most importantly, we can reconcile the two pictures by revisiting the definition of diffusional exchange. While the common perspective is that exchange entails passing through membranes, we advocate for the view of exchange as a process of homogenisation via transition between any regions that exhibit different diffusion processes. This more general notion is supported by recent work on the influence of geometry on filter-exchange-estimated exchange rates (Khateri et al., 2022) which showed that membrane permeation and transition between different domains of an anisotropic compartment bear similar imprints on the signal. Under this definition of exchange as "the temporal loss of observed diffusional heterogeneity," both membrane permeation and structural disorder are expected to have similar imprints on the signal variation with Γ. Regardless of which mechanism gives rise to apparent exchange, the observation that the signal variation across all 150 waveforms is smooth supports the view that the restriction- and exchange-weighting parameters, Γ and $V_\omega$, are sufficient to characterise both sources of diffusion time dependence in living tissue.

This work found that the cerebellar cortex exhibits faster exchange than any other part of the brain (Figs. 7-8, Table 2). Here, it is worth noting that the anatomy of the cerebellar cortex is highly complex, featuring three layers of different cellular makeup: the granular, Purkinje and molecular layers (Consalez et al., 2021; Nguyen et al., 2021; Tax et al., 2020; Voogd and Glickstein, 1998). The granular layer comprises the abundant granule cells with soma sizes of approximately 7 μm. The granule cells have unmyelinated axons that rise towards the molecular layer where they bifurcate into two branches building the so-called parallel fibres (Voogd and Glickstein, 1998). The Purkinje layer contains Purkinje cells which are 25-40 μm in diameter and have an intricate dendritic tree that intersects the parallel fibres. Axons of Purkinje cells are myelinated and descend towards the cerebellar nuclei and white matter. With this picture in mind, exchange in the cerebellar cortex may be occurring between the granule cells and their neurites or between neurites and the extracellular space. Alternatively, following the view of exchange as a homogenisation process, the observed exchange contrast may arise from diffusing water transitioning between different domains of the highly disordered extracellular environment of the cerebellar cortex. We note that considerable time dependence in the cerebellum has also been observed in animal models by studies using spectrally tuned gradient waveforms (Lundell et al., 2019) as well as combinations of SDE and OGSE acquisitions (Aggarwal et al., 2020; Wu et al., 2021, 2014). However, as noted earlier, the sources of time dependence in those studies cannot be disentangled without independently varying the restriction- and exchange-weighting. Note that with increasing oscillation frequency in OGSE acquisitions, both the restriction and exchange weighting increase leading to a reduction in the signal amplitude. Thus, one effect may be mistaken for the other in variable-frequency OGSE data.

A noteworthy outcome of this work is that exchange is considerably faster in grey matter than in white matter (Figs. 7-8, Table 2). Negligible exchange rates in healthy white matter at clinical diffusion times have been reported by others and explained by the presence of myelin (Nilsson et al., 2013b, 2013a, 2009). The results presented in this work suggest exchange times of 115 ms or longer in grey matter. This finding seems to be at odds with recent work suggesting exchange rates in grey matter of 15 ms and below (Jelescu et al., 2022; Olesen et al., 2022) but is in agreement with filter-exchange imaging studies that report exchange times in grey matter of a few 100 ms (Lampinen et al., 2017; Nilsson et al., 2013a). Overall, the reliable estimation of exchange rates *in vivo* with diffusion MRI is challenging due in part to shortcomings in both modelling and encoding strategies. Regarding the former, a prominent example is the Kärger model (Kärger, 1985). While this model is the mainstay of all exchange estimation in diffusion MRI, its assumptions are generally violated in recent literature reporting short exchange times in grey matter (Fieremans et al., 2010; Jelescu et al., 2022; Olesen et al., 2022). The notion that exchange and restricted diffusion exhibit an interplay has been held for a long time (Meier et al., 2003; Nilsson et al., 2013b; Stanisz et al., 1997) but has only recently gained traction in modelling (Jelescu et al., 2022; Jiang et al., 2022; Olesen et al., 2022). On the encoding front, we recently proposed a framework that leverages free waveforms to probe restricted diffusion and exchange, as an extension of conventional designs that are based on pulsed gradients (Chakwizira et al., 2022). As we demonstrated in our previous work, relying solely on pulsed gradients conflates the effects of restriction and exchange and creates degeneracies in model fitting. In the present work, we have shown, for the first time, that restriction and exchange can be separated into two independent contrasts *in vivo* and thus facilitate a more convincing inversion of the accompanying signal representation. While some bias is inevitable given the approximative nature of the signal equation, we claim that where the assumptions of the framework are respected, the obtained exchange estimates are unconfounded by restricted diffusion.

As our analysis indicated, ultra-strong gradients are beneficial for probing diffusion time dependence (Figs 7-8). This result is not surprising because stronger gradients enable a larger range of restriction- and exchange-weightings at high b-values in a relatively short total encoding time. Exchange rates observed in this study increased with the gradient strength because stronger gradients allowed protocols featuring lower minimum exchange-weighting times (Table 2, Table 1). We remark that the strong influence of the minimum available exchange-weighting suggests that exchange in tissue is likely multiexponential, which would imply—as observed in this work—that protocols with lower minimum exchange-weighting detect faster exchange regimes. However, despite the difference in absolute values, the grey-white matter contrast in the exchange maps is preserved across gradient strengths. On the contrary, there is less visual correspondence between restriction maps at 80 mT/m and those at 200 and 300 mT/m (Figs 7-8), indicating that the gain of using ultra-strong gradients is more pronounced for restricted diffusion than for exchange. This finding has a useful implication that is worth underlining: using free waveforms and the restriction-exchange framework applied in this study, grey-white matter exchange contrast in the human brain is obtainable even at 80 mT/m. In light of clinical feasibility, this study also showed that when combined with ultra-strong gradients, free waveforms can produce the same restriction and exchange contrasts in 4 minutes as would be obtained in 25 minutes (the duration of the full 300 mT/m application protocol). This is important for future clinical application where time is generally highly limited.

Some limitations of the current work are worth discussing. Regarding the parameter maps obtained with the application protocols (Figs 7-8), while nothing about the mean diffusivity and diffusivity variance maps is unusual, we note that the general lack of contrast in the restriction maps especially in the cerebrum is peculiar. One explanation is, as illustrated in Fig. 4A, that the dependence of $E_{\beta_2}$ on the fourth power of the cell diameter suppresses the contrast at small sizes. Another possible explanation is that the parameter $E_{\beta_2}$ is capturing extracellular rather than intracellular time dependence. As underlined in previous work, diffusion in the extracellular space has a linear dependence on frequency (Novikov et al., 2019) which is not accounted for by the restriction-weighting parameter $V_\omega$ that assumes a quadratic dependence. Regarding the exchange maps, it should be noted that the higher values observed in grey matter could result from crosstalk between large sizes and exchange, as illustrated using simulations in Fig. 4. However, a counterargument is that such crosstalk is associated with high values of the restriction coefficient (Fig. 4). It therefore cannot explain the grey-white matter contrast in the exchange maps because this study detected no contrast between the restriction coefficients in grey and white matter (Figs 7-8). Another important limitation of this study is that it neglects effects other than restriction and exchange that may confound the interpretation of the observed contrasts. Examples of such effects are anisotropy (Lasič et al., 2022; Lundell et al., 2019) and intra-compartmental kurtosis (Henriques et al., 2021, 2020).

In conclusion, this work has demonstrated that diffusion time dependence in the human brain is well-characterised by two parameters: one for restriction weighting and another for exchange weighting. By using gradient waveforms that independently vary these two parameters, it is possible to separate independent contrasts driven by restriction and exchange in different regions of the brain as well as obtain plausible exchange estimates especially in grey matter. Future work will address the shortcomings highlighted above and explore the potential of the framework for characterising brain tumours.

## Acknowledgments

The authors acknowledge the Master Research Agreement between Skåne University Hospital and Siemens Healthcare. This research was funded by: Massachusetts Life Science Foundation, VR (Swedish Research Council) (grant numbers 2016-03443, 2020-04549 and 2021-04844), eSSENCE (grant number 6:4), Cancerfonden (The Swedish Cancer Society) (grant numbers 2019/474 and 22 0592 JIA) and NIH (National Institutes of Health) (grant numbers R01NS125781, R01MH074794 and P41EB015902).

## Declarations of interest

Chakwizira A: none, Westin C-F: none, Szczepankiewicz F: none, Nilsson M: none. Zhu A and Foo T are employees of GE Research.

### Data and Code Availability

Data can be made available upon request subjective to institutes' policies.

# References

Aggarwal, M., Smith, M.D., Calabresi, P.A., 2020. Diffusion-time dependence of diffusional kurtosis in the mouse brain. Magn Reson Med 84, 1564–1578. https://doi.org/10.1002/mrm.28189

Alexander, D.C., Hubbard, P.L., Hall, M.G., Moore, E.A., Ptito, M., Parker, G.J.M., Dyrby, T.B., 2010. Orientationally invariant indices of axon diameter and density from diffusion MRI. NeuroImage 52, 1374–1389. https://doi.org/10.1016/j.neuroimage.2010.05.043

Callaghan, P.T., Coy, A., MacGowan, D., Packer, K.J., Zelaya, F.O., 1991. Diffraction-like effects in NMR diffusion studies of fluids in porous solids. Nature 351, 467–469. https://doi.org/10.1038/351467a0

Chakwizira, A., Westin, C.-F., Brabec, J., Lasič, S., Knutsson, L., Szczepankiewicz, F., Nilsson, M., 2022. Diffusion MRI with pulsed and free gradient waveforms: Effects of restricted diffusion and exchange. NMR in Biomedicine n/a, e4827. https://doi.org/10.1002/nbm.4827

Chronik, B.A., Rutt, B.K., 2001. Simple linear formulation for magnetostimulation specific to MRI gradient coils. Magnetic Resonance in Medicine 45, 916–919. https://doi.org/10.1002/mrm.1121

Consalez, G.G., Goldowitz, D., Casoni, F., Hawkes, R., 2021. Origins, Development, and Compartmentation of the Granule Cells of the Cerebellum. Frontiers in Neural Circuits 14.

Evans, T.C., Jehle, D., 1991. The red blood cell distribution width. The Journal of Emergency Medicine 9, 71–74. https://doi.org/10.1016/0736-4679(91)90592-4

Fan, Q., Eichner, C., Afzali, M., Mueller, L., Tax, C.M.W., Davids, M., Mahmutovic, M., Keil, B., Bilgic, B., Setsompop, K., Lee, H.-H., Tian, Q., Maffei, C., Ramos-Llordén, G., Nummenmaa, A., Witzel, T., Yendiki, A., Song, Y.-Q., Huang, C.-C., Lin, C.-P., Weiskopf, N., Anwander, A., Jones, D.K., Rosen, B.R., Wald, L.L., Huang, S.Y., 2022. Mapping the human connectome using diffusion MRI at 300 mT/m gradient strength: Methodological advances and scientific impact. Neuroimage 254, 118958. https://doi.org/10.1016/j.neuroimage.2022.118958

Fieremans, E., Burcaw, L.M., Lee, H.-H., Lemberskiy, G., Veraart, J., Novikov, D.S., 2016. In vivo observation and biophysical interpretation of time-dependent diffusion in human white matter. NeuroImage 129, 414–427. https://doi.org/10.1016/j.neuroimage.2016.01.018

Fieremans, E., Novikov, D.S., Jensen, J.H., Helpern, J.A., 2010. Monte Carlo study of a two-compartment exchange model of diffusion. NMR Biomed 23, 711–724. https://doi.org/10.1002/nbm.1577

Foo, T.K.F., Tan, E.T., Vermilyea, M.E., Hua, Y., Fiveland, E.W., Piel, J.E., Park, K., Ricci, J., Thompson, P.S., Graziani, D., Conte, G., Kagan, A., Bai, Y., Vasil, C., Tarasek, M., Yeo, D.T.B., Snell, F., Lee, D., Dean, A., DeMarco, J.K., Shih, R.Y., Hood, M.N., Chae, H., Ho, V.B., 2020. Highly efficient head-only magnetic field insert gradient coil for achieving simultaneous high gradient amplitude and slew rate at 3.0T (MAGNUS) for brain microstructure imaging. Magn Reson Med 83, 2356–2369. https://doi.org/10.1002/mrm.28087

Henriques, R.N., Jespersen, S.N., Shemesh, N., 2021. Evidence for microscopic kurtosis in neural tissue revealed by correlation tensor MRI. Magnetic Resonance in Medicine 86, 3111–3130. https://doi.org/10.1002/mrm.28938

Henriques, R.N., Jespersen, S.N., Shemesh, N., 2020. Correlation tensor magnetic resonance imaging. NeuroImage 211, 116605. https://doi.org/10.1016/j.neuroimage.2020.116605

Huang, S.Y., Tian, Q., Fan, Q., Witzel, T., Wichtmann, B., McNab, J.A., Bireley, J.D., Machado, N., Klawiter, E.C., Mekkaoui, C., Wald, L.L., Nummenmaa, A., 2020. High-gradient diffusion MRI reveals distinct estimates of axon diameter index within different white matter tracts in the in vivo human brain. Brain Struct Funct 225, 1277–1291. https://doi.org/10.1007/s00429-019-01961-2

Huang, S.Y., Witzel, T., Keil, B., Scholz, A., Davids, M., Dietz, P., Rummert, E., Ramb, R., Kirsch, J.E., Yendiki, A., Fan, Q., Tian, Q., Ramos-Llordén, G., Lee, H.-H., Nummenmaa, A., Bilgic, B., Setsompop, K., Wang, F., Avram, A.V., Komlosh, M., Benjamini, D., Magdoom, K.N., Pathak, S., Schneider, W., Novikov, D.S., Fieremans, E., Tounekti, S., Mekkaoui, C., Augustinack, J., Berger, D., Shapson-Coe, A., Lichtman, J., Basser, P.J., Wald, L.L., Rosen, B.R., 2021. Connectome 2.0: Developing the next-generation ultra-high gradient strength human MRI scanner for bridging studies of the micro-, meso- and macro-connectome. Neuroimage 243, 118530. https://doi.org/10.1016/j.neuroimage.2021.118530

Jelescu, I.O., de Skowronski, A., Geffroy, F., Palombo, M., Novikov, D.S., 2022. Neurite Exchange Imaging (NEXI): A minimal model of diffusion in gray matter with inter-compartment water exchange. NeuroImage 256, 119277. https://doi.org/10.1016/j.neuroimage.2022.119277

Jelescu, I.O., Palombo, M., Bagnato, F., Schilling, K.G., 2020. Challenges for biophysical modeling of microstructure. Journal of Neuroscience Methods 344, 108861. https://doi.org/10.1016/j.jneumeth.2020.108861

Jiang, X., Devan, S.P., Xie, J., Gore, J.C., Xu, J., 2022. Improving MR cell size imaging by inclusion of transcytolemmal water exchange. NMR Biomed 35, e4799. https://doi.org/10.1002/nbm.4799

Jiang, X., Devan, S.P., Xie, J., Gore, J.C., Xu, J., 2016. Improving MR cell size imaging by inclusion of transcytolemmal water exchange. NMR in Biomedicine n/a, e4799. https://doi.org/10.1002/nbm.4799

Jones, D.K., Alexander, D.C., Bowtell, R., Cercignani, M., Dell'Acqua, F., McHugh, D.J., Miller, K.L., Palombo, M., Parker, G.J.M., Rudrapatna, U.S., Tax, C.M.W., 2018. Microstructural imaging of the human brain with a "super-scanner": 10 key advantages of ultra-strong gradients for diffusion MRI. Neuroimage 182, 8–38. https://doi.org/10.1016/j.neuroimage.2018.05.047

Kärger, J., 1985. NMR self-diffusion studies in heterogeneous systems. Advances in Colloid and Interface Science 23, 129–148. https://doi.org/10.1016/0001-8686(85)80018-X

Khateri, M., Reisert, M., Sierra, A., Tohka, J., Kiselev, V.G., 2022. What does FEXI measure? NMR in Biomedicine 35, e4804. https://doi.org/10.1002/nbm.4804

Klein, S., Staring, M., Murphy, K., Viergever, M.A., Pluim, J.P.W., 2010. elastix: A Toolbox for Intensity-Based Medical Image Registration. IEEE Transactions on Medical Imaging 29, 196–205. https://doi.org/10.1109/TMI.2009.2035616

Lampinen, B., Szczepankiewicz, F., van Westen, D., Englund, E., C Sundgren, P., Lätt, J., Ståhlberg, F., Nilsson, M., 2017. Optimal experimental design for filter exchange imaging: Apparent exchange rate measurements in the healthy brain and in intracranial tumors. Magn Reson Med 77, 1104–1114. https://doi.org/10.1002/mrm.26195

Lasič, S., Yuldasheva, N., Szczepankiewicz, F., Nilsson, M., Dall'Armellina, E., Budde, M.D., Schneider, J.E., Teh, I., Lundell, H., 2022. Stay on the beat with tensor-valued encoding: time-dependent diffusion and cell size estimation in ex vivo heart. Frontiers in Physics.

Lee, H.-H., Fieremans, E., Novikov, D.S., 2018. What dominates the time dependence of diffusion transverse to axons: Intra- or extra-axonal water? Neuroimage 182, 500–510. https://doi.org/10.1016/j.neuroimage.2017.12.038

Lee, H.-H., Jespersen, S.N., Fieremans, E., Novikov, D.S., 2020a. The impact of realistic axonal shape on axon diameter estimation using diffusion MRI. NeuroImage 223, 117228. https://doi.org/10.1016/j.neuroimage.2020.117228

Lee, H.-H., Papaioannou, A., Novikov, D.S., Fieremans, E., 2020b. In vivo observation and biophysical interpretation of time-dependent diffusion in human cortical gray matter. Neuroimage 222, 117054. https://doi.org/10.1016/j.neuroimage.2020.117054

Lundell, H., Nilsson, M., Dyrby, T.B., Parker, G.J.M., Cristinacce, P.L.H., Zhou, F.-L., Topgaard, D., Lasič, S., 2019. Multidimensional diffusion MRI with spectrally modulated gradients reveals unprecedented microstructural detail. Sci Rep 9, 9026. https://doi.org/10.1038/s41598-019-45235-7

Meier, C., Dreher, W., Leibfritz, D., 2003. Diffusion in compartmental systems. I. A comparison of an analytical model with simulations. Magn Reson Med 50, 500–509. https://doi.org/10.1002/mrm.10557

Montagnana, M., Danese, E., 2016. Red cell distribution width and cancer. Ann Transl Med 4, 399. https://doi.org/10.21037/atm.2016.10.50

Nguyen, T.M., Thomas, L.A., Rhoades, J.L., Ricchi, I., Yuan, X.C., Sheridan, A., Hildebrand, D.G.C., Funke, J., Regehr, W.G., Lee, W.-C.A., 2021. Structured connectivity in the cerebellum enables noise-resilient pattern separation. https://doi.org/10.1101/2021.11.29.470455

Nilsson, M., Lasič, S., Drobnjak, I., Topgaard, D., Westin, C.-F., 2017. Resolution limit of cylinder diameter estimation by diffusion MRI: The impact of gradient waveform and orientation dispersion. NMR in Biomedicine 30, e3711. https://doi.org/10.1002/nbm.3711

Nilsson, M., Lätt, J., Nordh, E., Wirestam, R., Ståhlberg, F., Brockstedt, S., 2009. On the effects of a varied diffusion time in vivo: is the diffusion in white matter restricted? Magn Reson Imaging 27, 176–187. https://doi.org/10.1016/j.mri.2008.06.003

Nilsson, M., Lätt, J., Westen, D. van, Brockstedt, S., Lasič, S., Ståhlberg, F., Topgaard, D., 2013a. Noninvasive mapping of water diffusional exchange in the human brain using filter-exchange imaging. Magnetic Resonance in Medicine 69, 1572–1580. https://doi.org/10.1002/mrm.24395

Nilsson, M., Szczepankiewicz, F., Westen, D. van, Hansson, O., 2015. Extrapolation-Based References Improve Motion and Eddy-Current Correction of High B-Value DWI Data: Application in Parkinson's Disease Dementia. PLOS ONE 10, e0141825. https://doi.org/10.1371/journal.pone.0141825

Nilsson, M., van Westen, D., Ståhlberg, F., Sundgren, P.C., Lätt, J., 2013b. The role of tissue microstructure and water exchange in biophysical modelling of diffusion in white matter. MAGMA 26, 345–370. https://doi.org/10.1007/s10334-013-0371-x

Ning, L., Nilsson, M., Lasič, S., Westin, C.-F., Rathi, Y., 2018. Cumulant expansions for measuring water exchange using diffusion MRI. J Chem Phys 148, 074109. https://doi.org/10.1063/1.5014044

Novikov, D.S., 2021. The present and the future of microstructure MRI: From a paradigm shift to normal science. Journal of Neuroscience Methods 351, 108947. https://doi.org/10.1016/j.jneumeth.2020.108947

Novikov, D.S., Fieremans, E., Jespersen, S.N., Kiselev, V.G., 2019. Quantifying brain microstructure with diffusion MRI: Theory and parameter estimation. NMR in Biomedicine 32, e3998. https://doi.org/10.1002/nbm.3998

Novikov, D.S., Kiselev, V.G., Jespersen, S.N., 2018. On modeling. Magnetic Resonance in Medicine 79, 3172–3193. https://doi.org/10.1002/mrm.27101

Olesen, J.L., Østergaard, L., Shemesh, N., Jespersen, S.N., 2022. Diffusion time dependence, power-law scaling, and exchange in gray matter. NeuroImage 251, 118976. https://doi.org/10.1016/j.neuroimage.2022.118976

Papadopoulos, M.C., Verkman, A.S., 2012. Aquaporin 4 and neuromyelitis optica. Lancet Neurol 11, 535–544. https://doi.org/10.1016/S1474-4422(12)70133-3

Paquette, M., Eichner, C., Knösche, T.R., Anwander, A., 2020. Axon Diameter Measurements using Diffusion MRI are Infeasible. bioRxiv 2020.10.01.320507. https://doi.org/10.1101/2020.10.01.320507

Price-Jones, C., 1922. The diameters of red cells in pernicious anæmia and in anæmia following hæmorrhage. The Journal of Pathology and Bacteriology 25, 487–504. https://doi.org/10.1002/path.1700250410

Reynaud, O., 2017. Time-Dependent Diffusion MRI in Cancer: Tissue Modeling and Applications. Front. Phys. 5. https://doi.org/10.3389/fphy.2017.00058

Ruggiero, M.R., Baroni, S., Pezzana, S., Ferrante, G., Geninatti Crich, S., Aime, S., 2018. Evidence for the Role of Intracellular Water Lifetime as a Tumour Biomarker Obtained by In Vivo Field-Cycling Relaxometry. Angewandte Chemie International Edition 57, 7468–7472. https://doi.org/10.1002/anie.201713318

Setsompop, K., Kimmlingen, R., Eberlein, E., Witzel, T., Cohen-Adad, J., McNab, J.A., Keil, B., Tisdall, M.D., Hoecht, P., Dietz, P., Cauley, S.F., Tountcheva, V., Matschl, V., Lenz, V.H., Heberlein, K., Potthast, A., Thein, H., Van Horn, J., Toga, A., Schmitt, F., Lehne, D., Rosen, B.R., Wedeen, V., Wald, L.L., 2013. Pushing the limits of in vivo diffusion MRI for the Human Connectome Project. Neuroimage 80, 220–233. https://doi.org/10.1016/j.neuroimage.2013.05.078

Shashni, B., Ariyasu, S., Takeda, R., Suzuki, T., Shiina, S., Akimoto, K., Maeda, T., Aikawa, N., Abe, R., Osaki, T., Itoh, N., Aoki, S., 2018. Size-Based Differentiation of Cancer and Normal Cells by a Particle Size Analyzer Assisted by a Cell-Recognition PC Software. Biological and Pharmaceutical Bulletin 41, 487–503. https://doi.org/10.1248/bpb.b17-00776

Sønderby, C.K., Lundell, H.M., Søgaard, L.V., Dyrby, T.B., 2014. Apparent exchange rate imaging in anisotropic systems. Magnetic Resonance in Medicine 72, 756–762. https://doi.org/10.1002/mrm.24957

Stanisz, G.J., Wright, G.A., Henkelman, R.M., Szafer, A., 1997. An analytical model of restricted diffusion in bovine optic nerve. Magnetic Resonance in Medicine 37, 103–111. https://doi.org/10.1002/mrm.1910370115

Szczepankiewicz, F., Sjölund, J., Ståhlberg, F., Lätt, J., Nilsson, M., 2019. Tensor-valued diffusion encoding for diffusional variance decomposition (DIVIDE): Technical feasibility in clinical MRI systems. PLOS ONE 14, e0214238. https://doi.org/10.1371/journal.pone.0214238

Szczepankiewicz, F., Westin, C.-F., Nilsson, M., 2021. Gradient waveform design for tensor-valued encoding in diffusion MRI. Journal of Neuroscience Methods 348, 109007. https://doi.org/10.1016/j.jneumeth.2020.109007

Tax, C.M.W., Szczepankiewicz, F., Nilsson, M., Jones, D.K., 2020. The dot-compartment revealed? Diffusion MRI with ultra-strong gradients and spherical tensor encoding in the living human brain. NeuroImage 210, 116534. https://doi.org/10.1016/j.neuroimage.2020.116534

Veraart, J., Novikov, D.S., Christiaens, D., Ades-aron, B., Sijbers, J., Fieremans, E., 2016. Denoising of diffusion MRI using random matrix theory. NeuroImage 142, 394. https://doi.org/10.1016/j.neuroimage.2016.08.016

Veraart, J., Nunes, D., Rudrapatna, U., Fieremans, E., Jones, D.K., Novikov, D.S., Shemesh, N., 2020. Noninvasive quantification of axon radii using diffusion MRI. eLife 9, e49855. https://doi.org/10.7554/eLife.49855

Verkman, A.S., Hara-Chikuma, M., Papadopoulos, M.C., 2008. Aquaporins--new players in cancer biology. J Mol Med (Berl) 86, 523–529. https://doi.org/10.1007/s00109-008-0303-9

Verkman, A.S., Tradtrantip, L., Smith, A.J., Yao, X., 2017. Aquaporin Water Channels and Hydrocephalus. Pediatr Neurosurg 52, 409–416. https://doi.org/10.1159/000452168

Volles, M.J., Lee, S.-J., Rochet, J.-C., Shtilerman, M.D., Ding, T.T., Kessler, J.C., Lansbury, P.T., 2001. Vesicle Permeabilization by Protofibrillar α-Synuclein: Implications for the Pathogenesis and Treatment of Parkinson's Disease. Biochemistry 40, 7812–7819. https://doi.org/10.1021/bi0102398

Voogd, J., Glickstein, M., 1998. The anatomy of the cerebellum. Trends Neurosci 21, 370–375. https://doi.org/10.1016/s0166-2236(98)01318-6

Wu, D., Martin, L.J., Northington, F.J., Zhang, J., 2014. Oscillating gradient diffusion MRI reveals unique microstructural information in normal and hypoxia-ischemia injured mouse brains. Magnetic Resonance in Medicine 72, 1366–1374. https://doi.org/10.1002/mrm.25441

Wu, D., Zhang, Y., Cheng, B., Mori, S., Reeves, R.H., Gao, F.J., 2021. Time-dependent diffusion MRI probes cerebellar microstructural alterations in a mouse model of Down syndrome. Brain Communications 3, fcab062. https://doi.org/10.1093/braincomms/fcab062

Xu, J., Jiang, X., Devan, S.P., Arlinghaus, L.R., McKinley, E.T., Xie, J., Zu, Z., Wang, Q., Chakravarthy, A.B., Wang, Y., Gore, J.C., 2021. MRI-Cytometry: Mapping non-parametric cell size distributions using diffusion MRI. Magn Reson Med 85, 748–761. https://doi.org/10.1002/mrm.28454

Xu, J., Li, H., Harkins, K.D., Jiang, X., Xie, J., Kang, H., Does, M.D., Gore, J.C., 2014. Mapping mean axon diameter and axonal volume fraction by MRI using temporal diffusion spectroscopy. NeuroImage 103, 10–19. https://doi.org/10.1016/j.neuroimage.2014.09.006

Yadav, N.N., Stait-Gardner, T., Price, W.S., 2010. Hardware Considerations for Diffusion MRI, in: Jones, P., Derek K. (Ed.), Diffusion MRI: Theory, Methods, and Applications. Oxford University Press, p. 0. https://doi.org/10.1093/med/9780195369779.003.0011

Zhang, H., Hubbard, P.L., Parker, G.J.M., Alexander, D.C., 2011. Axon diameter mapping in the presence of orientation dispersion with diffusion MRI. Neuroimage 56, 1301–1315. https://doi.org/10.1016/j.neuroimage.2011.01.084

# Supplementary material

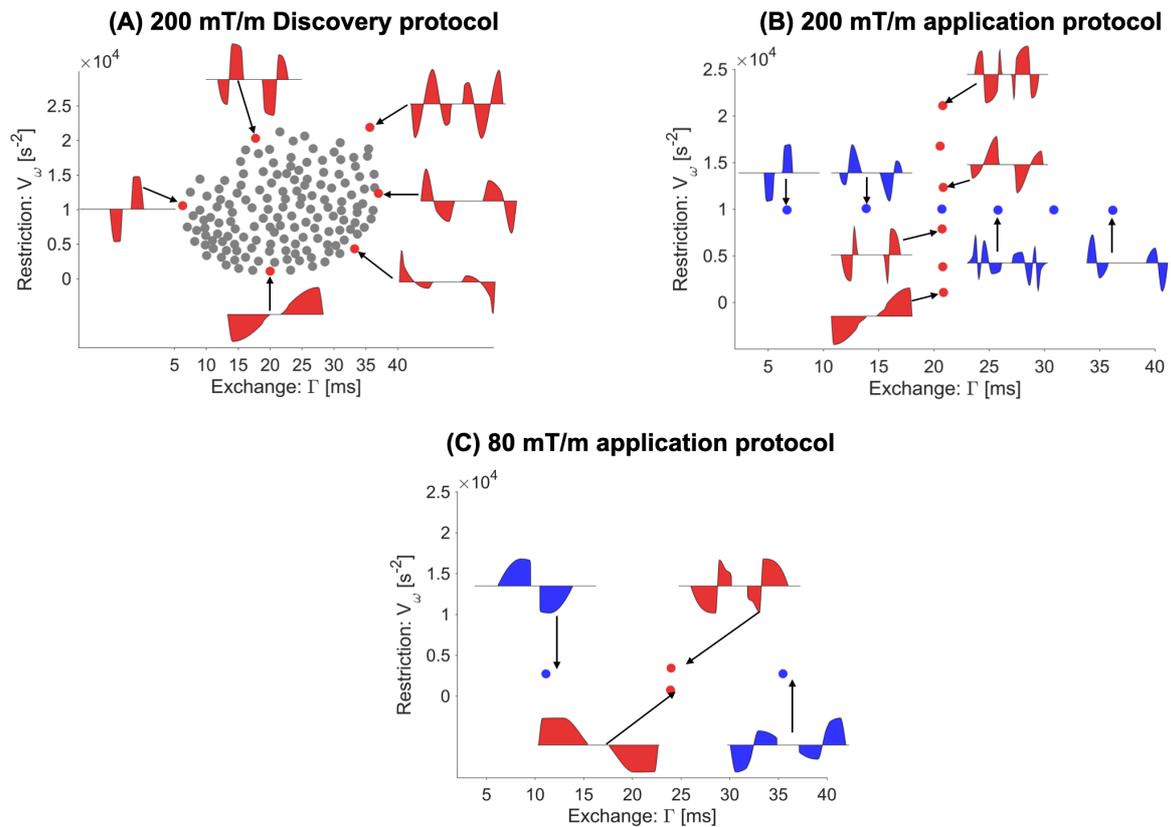

Figure A1: Protocols used for in vivo studies. (A) shows the 200 mT/m discovery protocol with a few waveform examples highlighted along the hull. (B) shows waveforms and restriction-exchange weightings of the 200 mT/m application protocol. (C) shows waveforms and restriction-exchange weightings for the 80 mT/m application protocol.

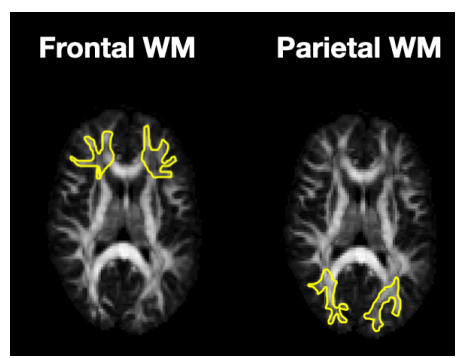

Figure A2: ROI placement in the frontal and parietal white matter.